\documentclass[twocolumn,showkeywords,amsmath,amssymb]{revtex4}

\usepackage{graphicx}
\usepackage{dcolumn}
\usepackage{bm}

\begin{document}

\title{Long-Distance Quantum Communication\\ with Entangled Photons using Satellites}

\author{Markus Aspelmeyer, Thomas Jennewein, and Anton Zeilinger}
 \affiliation{Institut f\"ur Experimentalphysik, Universit\"at Wien, Boltzmanngasse 5, A-1090 Wien, Austria}
\author{Martin Pfennigbauer and Walter Leeb}
 \affiliation{Institut f\"ur Nachrichtentechnik und
 Hochfrequenztechnik, Technische Universit\"at Wien,
 Gu\ss hausstra\ss e 25/389, A-1040 Wien, Austria}

\date{\today}

\begin{abstract}
The use of satellites to distribute entangled photon pairs (and
single photons) provides a unique solution for long-distance
quantum communication networks. This overcomes the principle
limitations of Earth-bound technology, i.e. the narrow range of
some 100~km provided by optical fiber and terrestrial free-space
links.
\end{abstract}

\maketitle

\section{\label{sec:Intro}Introduction}
Quantum entanglement is at the heart of quantum
physics~\cite{Schrodinger35a} and at the same time the basis of
most quantum communication protocols such as quantum
cryptography~\cite{Ekert91,Jennewein00,Tittel00,Naik00}, quantum
dense coding~\cite{Mattle96a}, quantum
teleportation~\cite{Bennett93a} or methods to exploit the
computational advantages of quantum communication
complexity~\cite{Brassard01,Buhrman99,Brukner02}. Each of those
schemes allows efficient communication and computation beyond the
capabilities of classical communication, which makes it
attractive as a new emerging quantum information technology. This
might lead to the build-up of a global quantum communication
network, where the distribution and manipulation of quantum
entanglement on a global scale is a central task. However, while
the realization of such schemes is routine work in the
laboratory, non-trivial problems emerge in long-distance
applications. At present, the only suitable system for
long-distance quantum communication are photons. Other systems
such as atoms or ions are studied thoroughly, however their
applicability for quantum communication schemes is presently not
feasible within the near future, leaving photons as the only
choice for long-distance quantum communication. One of the
problems of photon-based schemes is the loss of photons in the
quantum channel. This limits the bridgeable distance for single
photons to some estimated 100~km in present silica
fibers~\cite{Waks02,Gisin02}. In principle, this drawback can
eventually be overcome by subdividing the larger distance to be
bridged into smaller sections over which entanglement can be
teleported. The subsequent application of so-called ``entanglement
swapping''~\cite{Zukowski93a} may result in transporting of
entanglement over long distances. Additionally, to diminish
decoherence effects possibly induced by the quantum channel,
quantum purification might be applied to eventually implement a
full quantum repeater~\cite{Briegel98a}. In fact, the
experimental building blocks for a full-scale quantum repeater
based on linear optics have been successfully demonstrated over
the last years by the realization of teleportation and
entanglement swapping~\cite{Bouwmeester97a,Pan98a,Jennewein01}
and, only recently, by the implementation of a quantum
purification protocol~\cite{Pan03b}. Two related, recent results,
both of relevance for long-distance applications, are the
demonstration of quantum state teleportation over a distance of
several tens of meters~\cite{Marcikic03} and the first
realization of freely propagating teleported qubits~\cite{Pan03},
which eventually will allow the subsequent use of teleported
states. From the present point of view it seems obvious that a
full implementation of a quantum repeater is within reach.

Despite those achievements of quantum communication experiments,
the distances over which entanglement can be distributed in a
single section, i.e. without a quantum repeater in-between, are
by far not of a global scale. Experiments based on present fiber
technology have demonstrated that entangled photon pairs can be
separated by distances ranging from several hundreds of meters up
to 10~km~\cite{Tapster94,Tittel98,Weihs98}, but no significant
improvements are to be expected. On the other hand, optical
free-space links could provide a unique solution to this problem
since they allow in principle for much larger propagation
distances of photons due to the low absorption of the atmosphere
in certain wavelengths ranges. Also, the almost non-birefringent
character of the atmosphere~\cite{Buttler98c} guarantees the
preservation of polarization entanglement to a high degree.
Free-space optical links have been studied and successfully
implemented already for several years for their application in
quantum cryptography based on faint classical laser
pulses~\cite{Buttler98c,Hughes02,Kurtsiefer02}. A next crucial
step is the distribution of quantum entanglement via such free
space links.

However, terrestrial free-space links suffer from obstruction of
objects in the line of sight, from possible severe attenuation
due to weather conditions and aerosols~\cite{Horvath02} and,
eventually, from the Earth's curvature. They are thus limited to
rather short distances. To fully exploit the advantages of
free-space links, it will be necessary to use space and satellite
technology. By transmitting and/or receiving either photons or
entangled photon pairs to and/or from a satellite, entanglement
can be distributed over truly large distances and thus would
allow quantum communication applications on a global scale. Such
a scenario looks unrealistic at first sight, but in this paper we
will show that the demonstration of quantum communication
protocols using satellites is already feasible today. To do so,
we will describe possible space scenarios based on entanglement.
We then analyze prerequisites to distribute entanglement via
satellites, describe experimental scenarios for first
proof-of-principle experiments and finally give an outlook on the
perspectives of satellite-aided quantum communication.

\section{\label{sec:scenarios}Scenarios of Space Experiments}
When considering space scenarios that allow the distribution of
entangled photon pairs we can distinguish the cases in which a
satellite is used to carry either a transmitter of entangled
photons, or a receiver, or a relay station to distribute photons
to further locations. These scenarios will permit different
applications.

\subsection{Earth-based transmitter terminal}

\begin{figure}
\begin{center}
\includegraphics[width=1\columnwidth]{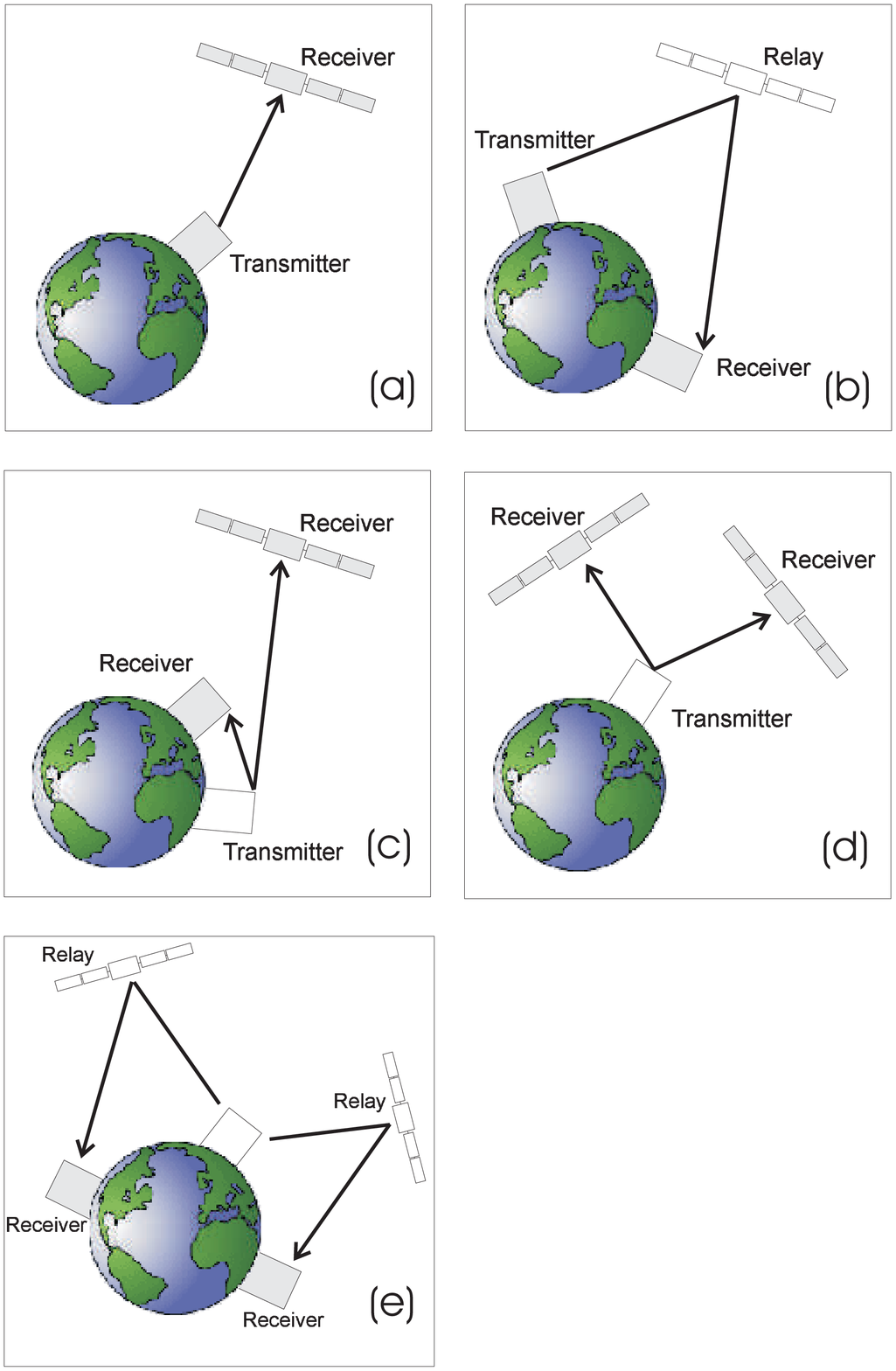}
\end{center}
\caption{Scenarios for satellite-aided quantum communication with
an Earth-based transmitter terminal. The transmitter terminal
distributes entangled photon pairs to the receivers which can
perform an entanglement-based quantum communication protocol. As
indicated, a relay module redirects and/or manipulates qubit
states without actually detecting
them.}\label{fig:ground_transmitter}
\end{figure}

The scenarios involving an Earth-based transmitter terminal allow
to share quantum entanglement between ground and satellite,
between two ground stations or between two satellites and thus to
communicate between such terminals employing quantum communication
protocols. In the most simple case, a straight uplink to one
satellite-based receiver (see
Fig.~\ref{fig:ground_transmitter}~a) can be used to perform
secure quantum key distribution between the transmitter station
and the receiver. Here, one of the photons of the entangled pair
is being detected right at the transmitter site and thus the
entangled photon source is used as a triggered source for single
photons. If the satellite acts as a relay station (see
Fig.~\ref{fig:ground_transmitter}~b), the same protocol can be
established between two distant Earth-based communication
parties. Shared entanglement between two parties can be achieved
by pointing each of the photons of an entangled pair either
towards an Earth-based station and a satellite or towards two
separate satellites (see Figs.~\ref{fig:ground_transmitter}~c and
d). Another set of satellite-based relays can be used to further
distribute the entangled photons to two ground stations (see
Fig.~\ref{fig:ground_transmitter}~e). Possible applications for
shared entanglement between two parties are quantum key
distribution or entanglement-enhanced communication
protocols~\cite{Bennett96c}.

\subsection{Space-based transmitter terminal}
\begin{figure}
\begin{center}
\includegraphics[width=1\columnwidth]{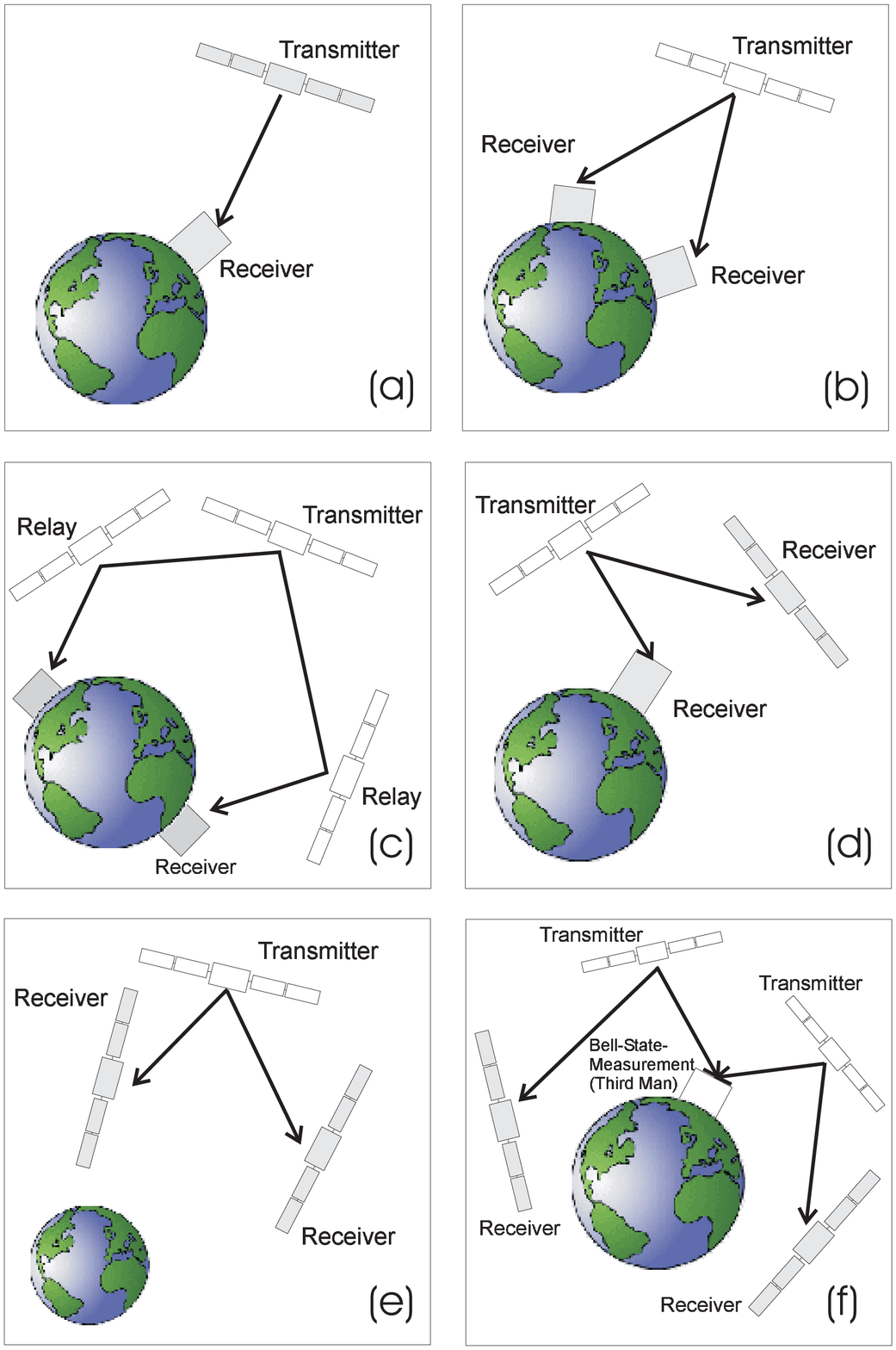}
\end{center}
\caption{Scenarios for quantum communication with a space-based
transmitter terminal. Since optical space-to-Earth downlinks are
less effected by atmospheric turbulence such configurations
provide lower overall link attenuations and thus allow longer
distances than the corresponding Earth-to-space
links.}\label{fig:space_transmitter}
\end{figure}

In a second scenario, a transmitter with an entangled photon
source is placed on a space-based platform. This allows not only
longer link distances because of reduced influence of atmospheric
turbulence (see Appendix~\ref{sec:linkAttenuation}). It will also
be the preferred configuration for global quantum communication,
since only one downlink per photon of the entangled pair is
necessary to share entanglement between two Earth-based
receivers. Again, already a simple downlink allows to establish a
single-photon link e.g. for quantum cryptography (see
Fig.~\ref{fig:space_transmitter}~a). In this configuration, a key
exchange between two ground stations is also possible. To this
end each of the two ground stations has to establish a quantum
key with the satellite. Since the space terminal has access to
both keys, it can transmit a logical combination of the keys,
which can then be used by either ground station or both ground
stations such that they arrive at the same key. This logical
combination can easily be chosen such that it cannot reveal any
information about the key. Note that the key does not have to be
generated simultaneously at both receiver stations. In principle,
a quantum key exchange can be performed between arbitrarily
located ground stations. This is also possible for a ground-based
transmitter terminal as shown in
Fig.~\ref{fig:ground_transmitter}~a. However, in all such
scenarios based on single photons the security requirement for
the transmitter terminal is as high as it is for the ground
station. Only the use of entangled states sent to two separate
ground stations allows instantaneous key exchange between these
two communicating earth-bound parties and also relaxes the
security requirement for the transmitter module. Furthermore,
more advanced quantum communication schemes will be feasible. The
required shared entanglement can be established either by two
direct downlinks (Fig.~\ref{fig:space_transmitter}~b) or by using
additional satellite relay stations
(Fig.~\ref{fig:space_transmitter}~c). Quantum entanglement can
also be distributed between a ground station and a satellite
(Fig.~\ref{fig:space_transmitter}~d) or between two satellites
(Fig.~\ref{fig:space_transmitter}~e). In an even more elaborate
scheme a third party might be involved, which is capable of
performing a Bell-state analysis on two independent photons
(Fig.~\ref{fig:space_transmitter}~f). This allows quantum state
teleportation and even entanglement swapping and could thus
resemble a large-scale quantum repeater for a truly global
quantum communication network. Applied to quantum cryptography,
this third party might be used to control the communication
between the two other parties in a "Third-Man" cryptography
protocol. For example in a polarization-based experiment,
depending on whether he performs a simple polarization analysis
of the independent photons or a Bell-state measurement, the
"Third-Man" can communicate secretly with either of the two
parties (or with both) or he can control whether the two can
communicate secretly or not without knowing the content of the
communication.

\subsection{Link requirements}
The maximal acceptable link attenuation for a quantum
communication system based on entangled photons is determined by
the timing resolution and the dark count rates of the detectors
used, as well as by the net production rate of the source. As the
minimum signal-to-noise ratio we assume that necessary for the
violation of a Bell-inequality (see
Appendix~\ref{appendix:link_req}). With a typical detection
efficiency $\eta_{Det}=0.3$, a photon production rate of
$P=5\cdot 10^5~s^{-1}$, an estimated total background count rate
of $S=10^3~s^{-1}$ and a coincidence timing window of $\Delta
\tau = 5\cdot 10^{-9}$~s, the link efficiency should obey
\begin{equation}
  \eta_{link} \geq 6.66...\cdot10^{-7}
\end{equation}
when following the calculations presented in
Appendix~\ref{appendix:link_req}. Roughly speaking, a total link
efficiency of $\eta_{link}=\eta_{link1} \eta_{link2} \approx
10^{-6}$ ($-$60~dB) is necessary.

The link attenuation is also important for determining the number
of photon pairs that can be received in a certain time window.
This could be crucial in scenarios where the links are only
available for short times as is for example the case in uplinks to
low orbiting LEO satellites.

\section{Link Attenuation}\label{GEO}
\begin{figure}
\begin{center}
\includegraphics[width=1\columnwidth]{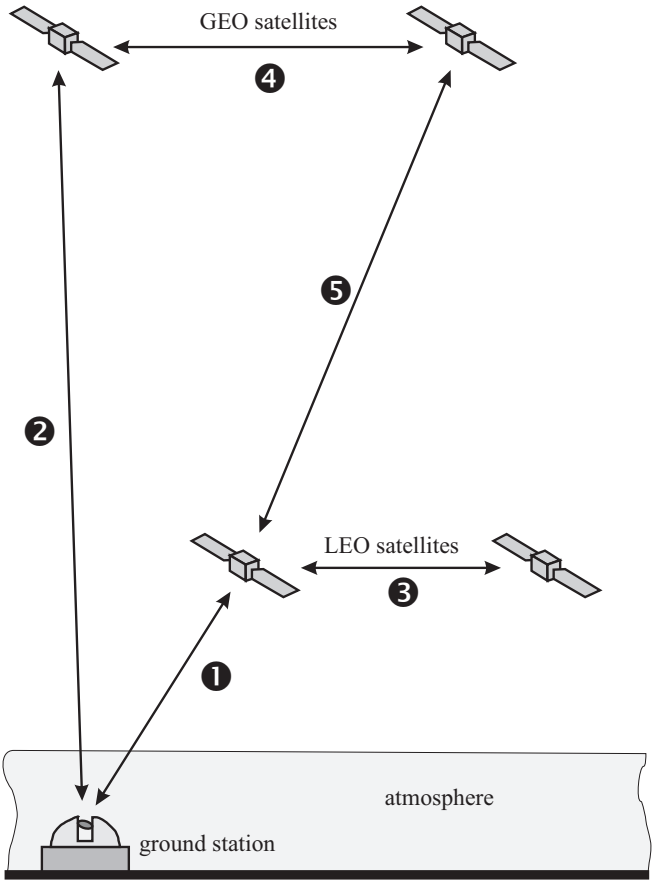}
\end{center}
\caption{Summary of links involving Earth-based ground stations
and LEO satellites or GEO satellites.}\label{fig:all_links}
\end{figure}

The overall link efficiency of 10$^{-6}$, corresponding to a
maximum attenuation of 60~dB, imposes quite a strong restriction
to the various space scenarios. In the following, we will
investigate the link attenuation for optical free-space links
involving space infrastructure. The attenuation factor calculated
includes the effect of beam diffraction, attenuation and
turbulence-induced beam spreading caused by the atmosphere,
receive aperture diameter, losses within the telescopes acting as
antennas, as well as antenna pointing loss. Effects not included
in this factor are the detection efficiency of the photon-counter
modules ($\sim$3.5 dB per detector) and reflective and absorptive
losses at optical components (typically $\sim$3 dB per individual
photon link). One may take these losses into account when
calculating individual link budgets. Figure~\ref{fig:all_links}
summarizes the scenarios considered based on satellites in
geostationary orbit (GEO) and in low earth orbit (LEO). Such
satellites may serve as a platform for transmitters or receivers.
We presently do not envision the use of passive relays, e.g.
retro-reflectors or mirrors, because of the high link loss they
would introduce and because of the difficulty to implement a
point-ahead angle~\footnote{The point-ahead angle denotes the
difference angle between transmit and receive direction of the
telescope. Its occurrence is a consequence of the movement of the
satellite together with the finite velocity of light of the signal
propagation.}

\subsection{Satellite--ground links}

\subsubsection{Ground--LEO or LEO--Ground links}
\begin{figure}[ht!]
            \begin{center}
                \includegraphics[width=1\columnwidth]{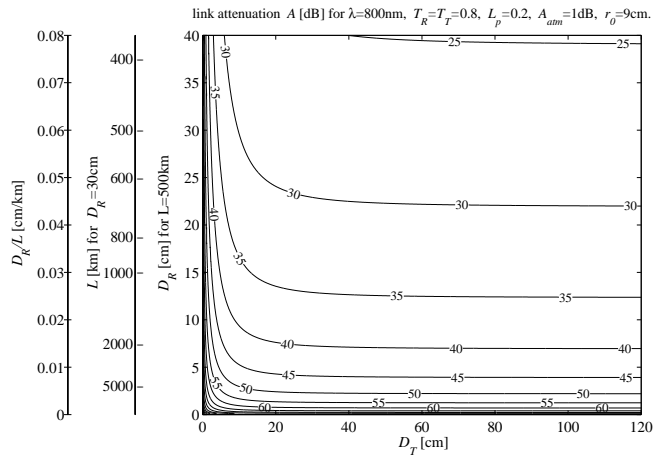}
                \caption{Contour plot of link attenuation $A$ (in dB) as a function of
                transmitter and receiver aperture diameter ($D_T$, $D_R$) and link
                distance $L$ for ground-to-LEO uplinks at $\lambda=800~nm$.}
                \label{fig:ground_leo_ul_mD_800}
            \end{center}
        \end{figure}

\begin{figure}[ht!]
            \begin{center}
                \includegraphics[width=1\columnwidth]{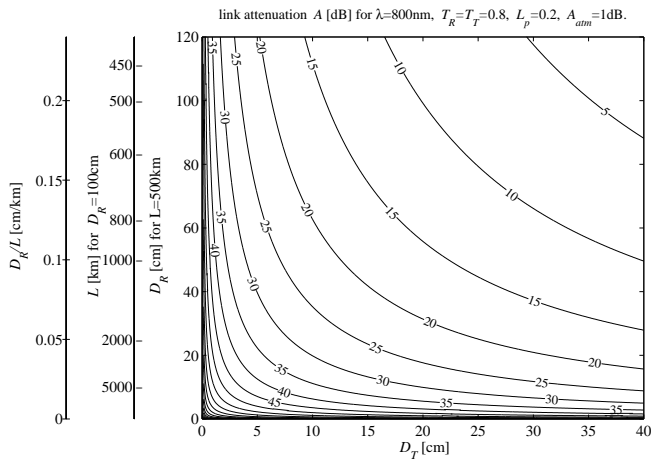}
                \caption{As Fig.~\ref{fig:ground_leo_ul_mD_800} but for LEO-to-ground downlinks.}
                \label{fig:ground_leo_dl_mD_800}
            \end{center}
        \end{figure}

For the case of a LEO-based transmitter or receiver (Link 1 in
Fig.~\ref{fig:all_links}), link attenuation poses no problems.
Even for quite small telescopes onboard the LEO satellite, the
attenuation factor is well below $60\,\mbox{dB}$ for all cases.
Figure~\ref{fig:ground_leo_ul_mD_800} is a contour plot of the
link attenuation as a function of transmitter and receiver
aperture diameter ($D_T$,$D_R$) for the ground-to-LEO uplinks
operated at a wavelength of $\lambda=800\,\mbox{nm}$. Two
additional vertical scales give the link distance L for
$30\,\mbox{cm}$ receive telescope aperture as well as for the
receive telescope aperture for a link distance of
L=$500\,\mbox{km}$ (For the equation and further parameters used
to arrive at Fig.~\ref{fig:ground_leo_ul_mD_800} - and also at
some subsequent figures - see
Appendix~\ref{sec:linkAttenuation}.). The lines of equal
attenuation are separated by $5\,\mbox{dB}$. The corresponding
plot for LEO-to-ground downlinks is shown in
Figure~\ref{fig:ground_leo_dl_mD_800}.
One notes that the
attenuation is much larger for the uplink than for the downlink.
This is caused by the pronounced influence of atmospheric
turbulence for the uplink, where the turbulent layers are close
to the transmitter. In contrast, for a downlink the effect of the
turbulent layer close to the receiver is negligible to first
order. Another consequence of turbulence is that increasing the
transmitter aperture for the uplink beyond $60\,\mbox{cm}$ hardly
decreases the link attenuation.

In the case of links connecting a LEO and a ground station the
possible duration for communication is comparatively short (e.g.\
a few minutes) and the angular velocity with which the telescope
at the ground station has to be moved to track the satellite
along its orbit is high. For all ground-to-space links the
possibility of communication is weather dependent. While for
clear weather and sufficient altitude of the ground station the
uplink attenuation caused by the atmosphere is mainly determined
by turbulence-induced beam spread~\footnote{For a diffraction
limited telescope of $20\,\mbox{cm}$ diameter and a link distance
of $500\,\mbox{km}$, some $20\,\mbox{dB}$ attenuation have to be
expected~\cite{pfennigbauer02_2}.}, clouded skies will make any
link impossible. This influence is augmented by the low elevation
angles typical for LEO-to-ground links and the thus increased
fraction of the propagation path within the atmosphere.

\subsubsection{Ground--GEO or GEO--Ground links}
The long distance in links between GEO and ground (Link 2 in
Fig.~\ref{fig:all_links}) results in a relatively high
attenuation. With a ground station aperture of $D=100$~cm and a
GEO terminal aperture of $D=30$~cm one will meet the 60~dB
requirement in a downlink, but not in an uplink (compare
Table~\ref{tab:all_links} of Section~\ref{sec:perspectives}).

\subsection{Satellite-satellite links\label{sec:sat-sat}}
While from a technological point of view a satellite-to-satellite
link is the most demanding configuration it offers highly
attractive scientific possibilities. It allows to cover, in
principle, arbitrarily large distances and might thus also be a
possibility for further novel fundamental tests on quantum
entanglement.

\begin{figure}
            \begin{center}
                \includegraphics[width=1\columnwidth]{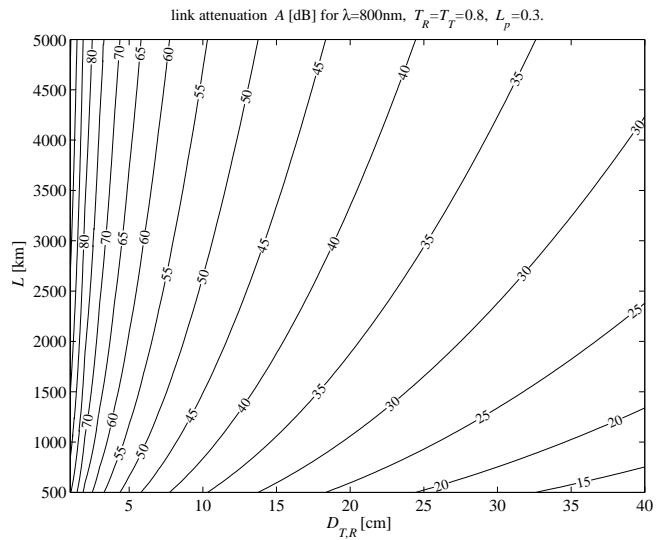}
                \caption{As Fig.~\ref{fig:ground_leo_ul_mD_800} but for LEO-to-LEO links.}
                \label{fig:leo_leo_LD_800}
            \end{center}
        \end{figure}

We calculated the attenuation factor as a function of transmitter
and receiver aperture diameter for a LEO-LEO link (Link 3 in
Fig.~\ref{fig:all_links}). Figure~\ \ref{fig:leo_leo_LD_800}
displays the attenuation for a wavelength of 800~nm as a function
of the satellite distance $L$ and telescope diameters $D_T$ and
$D_R$, assumed to be equal for both terminals. We conclude, that
the $60\,\mbox{dB}$-limit poses no problem for LEO-LEO links with
reasonable link distance.

\begin{figure}
            \begin{center}
                \includegraphics[width=1\columnwidth]{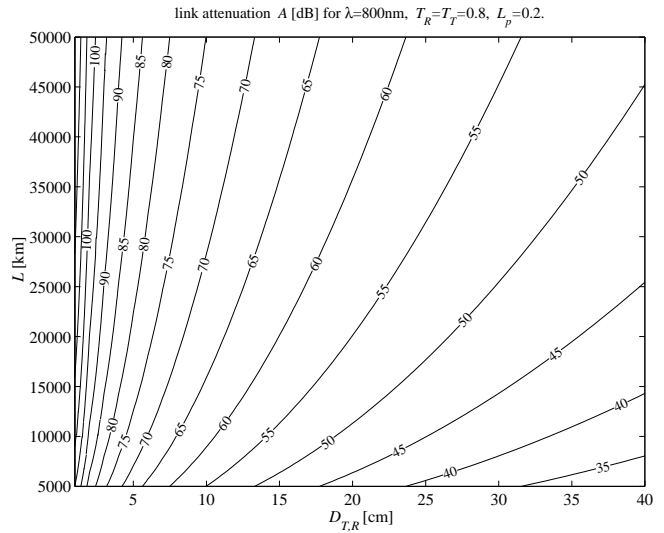}
                \caption{As Fig.~\ref{fig:ground_leo_ul_mD_800} but for GEO-to-GEO links.}
                \label{fig:geo_geo_LD_800}
            \end{center}
        \end{figure}

For GEO-GEO links (Link 4 in Fig.~\ref{fig:all_links}), the
attenuation can be read off Fig.\ \ref{fig:geo_geo_LD_800}, where
again equal telescope apertures have been assumed. For a distance
of $L=45000\,\mbox{km}$, an attenuation of $A=55$~dB would result
for $D_T=D_R=30$~cm.

\section{\label{sec:prereq}Technological Prerequisites for Entanglement in Space}

For performing the quantum communication experiments as described
above, the following minimum hardware is required: a transmitter
terminal to generate and transmit the entangled particles and one
or more receiver terminals suited for single-photon manipulation
and detection. Below we briefly outline their main features.

\subsection{Transmitter, receiver and relay modules}
A transmitter module comprises a photon source for entangled
photon pairs (including passive or active manipulation of single
qubit-states), a module for timing synchronization with the
receiver station and channel for classical communication. Present
entangled photon sources rely on laser-pumped spontaneous
parametric down conversion~\cite{Bouwmeester00}. This technology
is very likely capable of being miniaturized to a size suitable
for satellite modules.

A receiver module comprises one ore more optical input channels,
each of which allows independent manipulation of qubits such as
the rotation of photon polarization or the modulation of an
interferometric phase. Additionally, it has to be equipped with
single-photon detectors at each input port, a receiver module for
timing-synchronization, and a classical channel for communication
with the transmitter. Depending on whether active (remote)
control of optical elements for qubit manipulation is possible or
not (via, e.g., a polarizer or a retarder), we distinguish between
\textit{active} and \textit{passive} receiver modules. Passive
manipulation only requires a static setup of linear optics
components. Typically, beam splitters in the input ports would
randomly distribute incoming photons to differently oriented
retarders, polarizers or beam splitters, where a manipulation and
successive detection of single photons takes place. This kind of
passive receiver module has recently also been suggested by
Rarity~\textit{et~al.} in a space-suited system for single-photon
quantum key distribution~\cite{Rarity02_1}. For active control of
single qubit manipulation, an additional information concerning
the arrival time (i.e. a timing synchronization) is required.
More advanced quantum communication schemes, such as quantum
dense coding or quantum teleportation, require at some stage of
the protocol the projection of independent photons in a joint
Bell-state, i.e. a Bell-state measurement. Up to now, efficient
achievement of this projection is only possible by means of
two-particle-interferometry~\cite{Mattle96a}~\footnote{Recently,
another scheme based on the nonlinear couplings has been
suggested to achieve a full discrimination of all four Bell
states although with low efficiency~\cite{Kim01}}.This requires
that the arrival time difference of two photons at the two input
ports of a beam splitter has to be less than their coherence time
(typically 0.5~ps)~\cite{Zukowski95}.

A relay module would redirect and/or manipulate qubit states
without actually detecting them. Possibilities for its
implementation range from a simple retroreflector to a more
sophisticated relay-satellite (e.g. for deep-space
communications), where the entanglement of photons establishing
the quantum channel between ground station and deep-space
satellite could be purified. We emphasize, that a relay can not
serve as an amplifier. This is a consequence of the quantum
no-cloning theorem~\cite{Wooters82}.

\subsection{Existing optical space technology}
Optical transceivers for space-to-ground links or intersatellite
links are almost state of the art. The major design parameters for
the transmission sub-system are laser wavelength, modulation
format and data rate, and reception technique. Of equal
importance is the sub-system required for beam
\underline{p}ointing, link \underline{a}cquisition, and automatic
mutual terminal \underline{t}racking (PAT). Because of the very
narrow widths of the communication beams involved, PAT asks for
highly sophisticated concepts and for electro-mechanic and
electro-optic hardware meeting exceptional technological
standards. Major parameters entering the link capacity are
telescope size, optical transmit power, link distance, and
receiver sensitivity. Other aspects are mass, volume, and power
consumption of the terminal. Examples for existing space laser
communication links include ESA's intersatellite link SILEX
(Semiconductor Laser Intersatellite Link Experiment) and a
satellite ground link, which was only recently realized between
ARTEMIS and ESA's optical ground station OGS at Tenerife.

Photon sources and detectors presently implemented in such
classical space laser communication systems  can, in general, not
be directly employed in quantum communication systems. However,
the experience available may serve as a starting point for the
development of space qualified components needed for quantum
space experiments. The available optical communication technology
could of course be applied to provide the classical channel that
is always necessary in parallel to the quantum channel. One would
also make synergistic use of some of the optics employed for PAT
and employ one and the same telescope as antenna for both the
classical and the quantum channel, which is a novel way of
quantum-classical-multiplexing.

\subsection{\label{sec:principles}Proof-of-principle Experiments}
The establishment of entanglement in space and, subsequently, its
use for fundamental quantum physics experiments and quantum
communication applications necessitates certain experimental
stages:
\subsubsection*{\textbf{Stage I}}
Creation and detection of qubits (here: single photons) via
an optical space link. From an application point of view this
achievement would already allow to perform quantum key
distribution based on single photons.
\subsubsection*{\textbf{Stage II}}
Establishment of entanglement (i.e. non-classical
correlations via shared entangled particles) between the
communicating parties. This includes the ability to detect single
qubits synchronously at the spatially separated locations of the
communicating parties. This stage already allows the most
fundamental experiment in quantum physics, the violation of
Bell's inequality. It also makes possible further experiments such
as quantum key distribution based on entangled qubits.
\subsubsection*{\textbf{Stage III}}
Bell-state analysis of independent qubits.
For the case of photons, the most efficient scheme relies on
two-photon interferometry at a beam splitter. Technically
speaking, the arrival time of the photonic qubits at the receiver
module has to be synchronized such that photon wave-packets
overlap at the beam splitter within their coherence length. If
this problem is solved, all advanced quantum communication and
computation protocols such as quantum state teleportation or
quantum dense coding can be implemented.

The selection of an experimental scenario, in which all of these
stages can be performed, requires a trade-off between link
attenuation and experimental flexibility. It has been shown above
that the total link attenuation of the experimental setup must not
exceed some $60\,\mbox{dB}$, assuming present-day quantum-optics
technology. Presently, without the use of quantum memories,
entanglement can only be shared when more than one link is
available. For the symmetric case of two equally long quantum
links (one transmitter and two receivers), this limits the
maximal single-link attenuation (between one transmitter and one
receiver) to approx. $30\,\mbox{dB}$. Although space-to-space
links have the attractive advantage of not being influenced by
the Earth's atmosphere, we have to discard them at present due to
the expected disproportionate technological and financial effort
as compared to alternative schemes with at least one of the
communication terminals is on ground. Since two links should be
established, it is therefore most reasonable to place the
transmitter module in space, while the receiver modules stay in
easily accessible ground-based laboratories. Most envisioned
quantum experiments require higher flexibility at the receiver
due to active polarization control or data analysis. Also, the
atmosphere causes a larger footprint in an uplink than in a
downlink, due to the higher influence of turbulence. With such
perspectives it becomes obvious that in a first
proof-of-principle experiment one should place the transmitter
module into space and the receiver module(s) on Earth. Because of
their relative stationarity, terminals placed on GEO satellites
do not require such a highly sophisticated pointing, acquisition
and tracking (PAT) systems as those on a LEO satellite. They
would also allow for long-duration experiments. On the other
hand, the link attenuation and cost are significantly larger for
GEO links compared to LEO links. Therefore, when trading
GEO-based against LEO-based systems, we would rather accept the
more complex PAT system and the limited connection time per orbit
and suggest to use a LEO platform for the transmitter terminal
for first proof-of-principle experiments.

\section{\label{sec:perspectives}Perspectives and limits of space-aided quantum
communications} We determined the typical attenuation for the
links indicated in Fig.~\ref{fig:all_links}. The values listed in
Table~\ref{tab:all_links} were calculated using
equ.~\ref{equ:att_atm} for wavelengths of 800~nm and of 1550~nm,
respectively, assuming telescope apertures characteristic for
present day optical space technology. The first wavelength seems
reasonable because the best single-photon detectors exist for
800~nm, while the second wavelength is mainly used in standard
telecom systems. Applying this wavelength may thus increase
compatibility and reduce development effort because commercial
off-the-shelf components are available. However, the link
attenuation is slightly increased for $\lambda=$1550~nm due to
the higher absorption in the atmosphere and due to the higher
beam divergence at larger wavelength. For each link the default
parameters are specified in Appendix~\ref{appendix:parameters}.

Based on present-day technology and assuming reasonable link
parameters, it seems feasible to achieve enough entangled photons
per receiver pair to demonstrate a quantum communication protocol.
For example, assuming a LEO-based transmitter terminal, a
simultaneous link to two separate receiving ground stations and a
(conservatively estimated) total link attenuation of approx.\
$51\,\mbox{dB}$\footnote{Note again, that in times without
quantum memories, quantum entanglement can only be shared when
more than one link is available. For the symmetric case, i.e. two
equally long quantum links involving one transmitter and two
receivers, this limits the maximal single link attenuation
between one transmitter and one receiver to approx. 30~dB. Here we
assume a loss of $25.5\,\mbox{dB}$ for each of the downlinks.},
one can expect a local count rate of approx.\ $2600$ per second
in total at each of the receiver terminals. The number of shared
entangled photon pairs is then expected to be approx.\ $4$ per
second. For a link duration of 300~seconds this accumulates to a
net reception of 1200~entangled qubits. One can expect erroneous
detection events on the order of 7 per 100 seconds, which yields
a bit error of approx.\ 2\%. This would already allow a quantum
key distribution protocol between the two receiver stations. It is
thus clear, that a demonstration of basic quantum communication
protocols based on quantum entanglement can already be achieved
today.

  \begin{table}[ht!]
      \caption{Link attenuation for various space scenarios}
         \label{tab:all_links}
      \begin{center}
         \begin{tabular}{|c||c|c|c|}\hline
            \begin{tabular}{c}$800\,\mbox{nm}$ \\ \fbox{$1550\,\mbox{nm}$}\end{tabular} &
            \begin{tabular}{c}ground-based\\receiver\end{tabular} &
            \begin{tabular}{c}LEO\\receiver\end{tabular} &
            \begin{tabular}{c}GEO\\receiver\end{tabular} 
            \\\hline\hline
            \begin{tabular}{c}ground\\based\\transmitter\end{tabular}
            & &
            \begin{tabular}{c}$27.4\,\mbox{dB}$ \\ \fbox{$26.3\,\mbox{dB}$}\end{tabular} &
            \begin{tabular}{c}$64.5\,\mbox{dB}$ \\ \fbox{$63.4\,\mbox{dB}$}\end{tabular} 
            \\\hline
            \begin{tabular}{c}LEO\\transmitter\\ \end{tabular} &
            \begin{tabular}{c}$6.4\,\mbox{dB}$ \\ \fbox{$12.2\,\mbox{dB}$}\\\end{tabular} &
            \begin{tabular}{c}$28.5\,\mbox{dB}$ \\ \fbox{$33.6\,\mbox{dB}$}\\\end{tabular} &
            \begin{tabular}{c}$52.9\,\mbox{dB}$ \\\fbox{$58.6\,\mbox{dB}$}\\\end{tabular} 
            \\\hline
            \begin{tabular}{c}GEO\\transmitter\\ \end{tabular} &
            \begin{tabular}{c}$43.6\,\mbox{dB}$ \\ \fbox{$49.3\,\mbox{dB}$}\\\end{tabular} &
            \begin{tabular}{c}$52.9\,\mbox{dB}$ \\ \fbox{$58.6\,\mbox{dB}$}\\\end{tabular} &
            \begin{tabular}{c}$53.9\,\mbox{dB}$ \\\fbox{$59.7\,\mbox{dB}$}\\\end{tabular} 
            \\\hline

         \end{tabular}

      \end{center}
   For the numerical calculation
         the default parameter values given in Appendix~\ref{appendix:parameters} have been taken.
         The values within the boxes correspond to a wavelength of $1550\,\mbox{nm}$, while
         the others stand for $800\,\mbox{nm}$

         \end{table}

All proposed setups are based on the utilization of entangled
photon pairs as carrier of quantum information. Given
state-of-the-art technologies present in today's quantum optics
labs we can specify some practical limitations for the
preparation and detection of entangled qubits.

The rate of information transfer is limited by the maximal number
of photons or entangled photon pairs that can be created and
detected. Typical standard repetition rates for pulsed laser
sources able to create (entangled) qubit states are in the order
of $10^{6}- 10^{7}~s^{-1}$, which will faint out to only a few
thousands per second due to optical filtering, finite coupling
efficiency and finite detection efficiency. Additionally one has
to take into account transmission losses, which can limit the
qubit transmission drastically. Also, state of the art detector
systems have low dynamic ranges over a maximum of six orders of
magnitude~\footnote{The highest dynamic range of about six orders
of magnitude is achieved with Si avalanche photodiodes (APDs),
whereas systems based on InGaAs up to now reach only three to
four orders of magnitude.} and a maximal detection rate of some
MHz~\footnote{Some tens of kHz in the case of InGaAs.}. Further
development of source and detector technology will lead to
additional improvements of the qubit rates.

Besides the possibility of establishing a truly global quantum
communication network, space-based distribution of quantum
entanglement provides us with additional advantages. On the one
hand, quantum communication provides means to establish secure
and efficient communication. Space communication links certainly
match the category of links that should be secure and efficient
for several reasons: at first, satellite remote control is a
highly sensitive area with respect to security and, up to now, an
unsolved technical problem. Secondly, earth-to-satellite
communication not only requires considerable expense but is also
only possible within limited time-intervals (e.g. in the case of
LEO-satellites). Secondly, resources for communication are rare
and/or expensive, since active communication segments in space are
specified for low power consumption due to limited power
resources; therefore, their potential of communication to other
(space- or ground-segments) is strictly limited. Deep space
communication is a specific example where communication time is
limited and resources are expensive and efficient communication
is mandatory. Quantum communication complexity could provide a
unique means of information transport with very little
consumption of communication resources.

On the other hand, the distribution of quantum entanglement will
allow to expand the scale for testing the validity of quantum
physics by several orders of magnitude. This is a major challenge
for future fundamental experiments in quantum physics.

\newpage
\appendix
\section{Link requirements\label{appendix:link_req}}
The accidental coincidence rate is given by
\begin{equation}
C_{acc} = S_1 S_2 \Delta\tau,
\end{equation}
where $S_1, S_2$ are the dark count rates of the two detectors
and $\Delta\tau$ is the timing resolution for the electronic
registration of a two-fold coincidence event. As the minimum
signal to noise ratio we assume that required for violating a
Bell-inequality, since this guarantees at the same time the
security of certain quantum cryptography schemes~\cite{Fuchs97}.
For the case of polarization-entangled photons this necessitates
a two-fold coincidence visibility of at least 71\%, corresponding
to a signal-to-noise ratio~(SNR) of $6:1$~\footnote{The
visibility, in terms of the signal $S$ and the noise $N$, is
defined by $(S-N)/(S+N)$.}. Below that ratio a local realistic
modeling of the observed correlations is possible thus allowing
unobserved eavesdropping~\footnote{Note, that phase-coded
entanglement results in slightly higher requirements to show that
no local realistic model describes the corresponding
correlations~\cite{Aerts99}.}. Therefore,  in order to
discriminate the signal from the background coincidences, the
minimal observed coincidence rate $C_{min}$ must be at least 6
times larger than $C_{acc}$.

The coincidence detection rate is determined by the total
coincidence efficiency $\eta_{link}$, which is the product of the
individual efficiencies for the two qubit links,
\begin{equation}
\eta_{link}= \eta_{Link1}  \eta_{Link2} .
\end{equation}
The detected signal coincidences $C$ are given by the product
\begin{equation}
  C = P \eta_{link} \eta_{det1} \eta_{det2},
\end{equation}
where $P$ is the pair production rate of the source and
$\eta_{det1},\eta_{det2}$ are the detection probabilities. In
order to achieve a violation of Bell's inequality, the signal
coincidences must exceed the limit $C_{min}=SNR \cdot C_{acc}$,
which leads to the following limit for the total link efficiency
\begin{equation} \label{etalimit}
\eta_{link}\geq~ SNR \frac{C_{acc}}{P \eta_{det1}\eta_{det2}}=SNR
\frac{S_1 S_2 \Delta\tau}{P \eta_{det1}\eta_{det2}}.
\end{equation}

\section{Modeling the link attenuation\label{sec:linkAttenuation}}

        We define the link attenuation factor $A$ as the ratio of the mean transmit and receive
        power, $P_T$ and $P_R$~\cite{reyes02_2}, measured at the entrance and the exit of the
        transmit and the receive telescope, respectively. Thus losses due to single
        photon detection efficiency and optical elements such as filters, polarizers or
        retarders are not included in this number. Then the attenuation factor $A$ of a
        one-way free-space link is thus given by
        \begin{equation}
            A=\frac{L^2 \lambda^2}{D_T^2 D_R^2}\frac{1}{T_T(1-L_P)T_R}, \label{equ:att_stand}
        \end{equation}
        where $L$ is the link distance, $\lambda$ the wavelength, and
        $D_T$  and $D_R$ the diameters of the transmit and receive
        telescope. With $T_T$ and $T_R$ we denote the transmission factors
        ($\leq 1$) of the telescopes, $L_P$ is the pointing loss due
        misalignment of transmitter and receiver.
        This basic relationship applies (i)~if the receiver is in the transmitter's far
        field, i.e.\ $L\geq D_T^2/\lambda$, (ii)~if the transmit telescope is diffraction
        limited, and (iii)~if there is no influence of the atmosphere.

\subsection*{Influence of atmosphere\label{sec:atmosphere}}

Atmospheric effects on propagation at optical beams can be divided
into three categories: absorption, scattering, and
turbulence~\cite{winick86_1,andrews98_1}. While absorption and
scattering mainly depend on wavelength and visibility conditions,
the net impact of atmospheric turbulence additionally depends on
elevation angle and direction of transmission~\cite{fried66_1}.
The main effect of atmospheric turbulence is an enlarged beam
divergence, resulting in a reduced amount of signal power
collected by the receive telescope. Further turbulence-induced
effects are beam-wander, loss of coherence, scintillation and
pulse distortion and broadening~\footnote{Turbulence induced
pulse distortion and broadening might actually impose an upper
limit to the spectral bandwidth in the pulsed downconversion
schemes.}~\cite{Gilbert00}. The effect of turbulence is in general
quite different for a space-to-ground link and a ground-to-space
link. In a space-to-ground link the light propagates through
vacuum for the most of the distance first before being disturbed
by the atmosphere, whereas for a ground-to-space link the beam
spreading effects of turbulence take place at the beginning of the
propagation, causing a strongly enhanced divergence.


    \subsection*{Ground-to-space links}

        For ground-to-space links we therefore modify Equ.~\ref{equ:att_stand} to
        take into account an additional attenuation of the atmosphere
        and the influence of turbulence. The diffraction-limited divergence
        caused by the aperture diameter of the transmit telescope is
        increased when the beam passes turbulent atmosphere. The influence of the atmosphere
        can be taken into account by the so-called \emph{Fried parameter}, $r_0$, which
        can be interpreted as an ``effective aperture'' \cite{fried66_1}.
        We will assume that the divergence due to turbulence adds quadratically to the
        divergence of the telescope~\cite{pfennigbauer02_2}. The attenuation factor
        may then be approximated by
        \begin{equation}
            A=\frac{L^2 (\theta_T^2+\theta_{\text{atm}}^2)}{D_R^2}\frac{1}{T_T(1-L_P)T_R}10^{A_{\text{atm}}/10}, \label{equ:att_atm}
        \end{equation}
        where $A_{\text{atm}}$ is the attenuation of the atmosphere, given in dB.
        The divergence angle resulting from the transmit telescope is assumed to be
        \begin{equation}
            \theta_T=\frac{\lambda}{D_T} \label{equ:div_tel}
        \end{equation}
        and the turbulence causes the additional divergence
        \begin{equation}
            \theta_{\text{atm}}=\frac{\lambda}{r_0}. \label{equ:div_atm}
        \end{equation}

This calculation probably underestimates the turbulence
effect~\footnote{In comparison to the values presented
in~\cite{sodnik00_1}, the divergence obtained with our model is
lower by a factor of $1.5$. However, we do not know the exact
turbulence conditions assumed in~\cite{sodnik00_1}. Also, the
experimental results of the ARTEMIS--OGS downlink are slightly
worse than our calculations would predict.}, but our model is
considered to be suitable to calculate a lower-bound estimation
for the attenuation factor.

\section{Default parameters}\label{appendix:parameters}

For the altitude of the LEO satellite we assume 500~km, which
thus represents the lower limit of the link distance. For an
elevation angle of e.g. 15$^{\circ}$, the link distance is of some
1400~km~\footnote{Note that for low elevation angles the
influence of the atmosphere is increased, a fact not taken into
account by the model used here.}.

Geostationary satellites have an altitude of 36~000~km. Again, the
link distance may be larger, depending on the elevation of the
satellite (for the ARTEMIS-OGS link the link distance is
41~229~km).

The baseline for the ground aperture is 1~m because this is the
telescope diameter of ESA's optical ground station at Tenerife
(OGS). Telescopes with a diameter of 20 to 30~cm are small and
light enough to be operated even onboard a small LEO satellite.
Larger telescopes are feasible, especially for GEO satellites.

We assume the transmission factors $T_P$, $T_R$ of the involved
telescopes to be $0.8$.
The pointing loss is $L_P=0.2$ for all links except for the
LEO-LEO link, where we assume $L_P=0.3$ to take into account the
-- possibly -- high relative velocity of the satellites that
might result in reduced tracking accuracy.

The assumption of an atmospheric attenuation of $A_{atm}=1$~dB
applies for excellent sight conditions (no haze, fog, or clouds)
and is valid only in certain wavelength regions.

A recently obtained estimate for the Fried parameter, valid for
the optical ground station at Tenerife (OGS), is $r_0=90$~mm for
a wavelength of 800~nm in case of weak
turbulence~\cite{pfennigbauer02_2}.

For the calculations presented we have assumed a wavelength of
800~nm and 1550~nm. The following tables summarize the link
characteristics.

\subsection*{Table A I: Parameters for ground -- LEO and LEO -- ground links (default values are
underlined)}

        \begin{center}
            \begin{tabular}{|l|c|l|}\hline
                link distance           & $L$              & \underline{500}$\;\mbox{to}\;1400\,\mbox{km}$\\\hline
                ground aperture         & $D_T, D_R$       & $1\,\mbox{m}$                        \\\hline
                LEO aperture            & $D_T, D_R$       & $20\;\mbox{to}\;30\,\mbox{cm}$                   \\\hline
                wavelength              & $\lambda$        & $800\,\mbox{nm}$, $1550\,\mbox{nm}$ \\\hline
                telescope transmission factor     & $T_T=T_R$        & $0.8$                                \\\hline
                pointing loss           & $L_p$            & $0.2$                                \\\hline
                atmospheric attenuation & $A_{\text{atm}}$ & $1\,\mbox{dB}$                      \\\hline
                Fried parameter         & $r_0$            & $9\,\mbox{cm}$                      \\\hline
            \end{tabular}
        \end{center}
\subsection*{Table A II: Parameters for Ground -- GEO and GEO -- Ground links}

        \begin{center}
            \begin{tabular}{|l|c|l|}\hline
                link distance           & $L$              & $\geq 36~000\,\mbox{km}$                     \\\hline
                ground aperture         & $D_T, D_R$        & $1\,\mbox{m}$                        \\\hline
                LEO aperture            & $D_T, D_R$        & $20\;\mbox{to}\;30\,\mbox{cm}$                   \\\hline
                wavelength              & $\lambda$        & $800\,\mbox{nm}$, $1550\,\mbox{nm}$ \\\hline
                telescope transmission factor     & $T_T=T_R$        & $0.8$                                \\\hline
                pointing loss           & $L_p$            & $0.2$                                \\\hline
                atmospheric attenuation & $A_{\text{atm}}$ & $1\,\mbox{dB}$                      \\\hline
                Fried parameter         & $r_0$ & $9\,\mbox{cm}$                      \\\hline
            \end{tabular}
        \end{center}
\subsection*{Table A III: Parameters for LEO -- LEO links}

        \begin{center}
            \begin{tabular}{|l|c|l|}\hline
                link distance       & $L$       & $2~000\,\mbox{km}$                    \\\hline
                aperture            & $D_T=D_R$ & $20\;\mbox{to}\;30\,\mbox{cm}$                   \\\hline
                wavelength          & $\lambda$ & $800\,\mbox{nm}$, $1550\,\mbox{nm}$ \\\hline
                telescope transmission factor & $T_T=T_R$ & $0.8$                                \\\hline
                pointing loss       & $L_p$     & $0.3$                                \\\hline
            \end{tabular}
        \end{center}
\subsection*{Table A IV: Parameters for LEO -- GEO links}

        \begin{center}
            \begin{tabular}{|l|c|l|}\hline
                link distance       & $L$       & $35~500\,\mbox{km}$                  \\\hline
                aperture            & $D_T=D_R$ & $20\;\mbox{to}\;30\,\mbox{cm}$      \\\hline
                wavelength          & $\lambda$ & $800\,\mbox{nm}$, $1550\,\mbox{nm}$ \\\hline
                telescope transmission factor & $T_T=T_R$ & $0.8$                               \\\hline
                pointing loss       & $L_p$     & $0.2$                               \\\hline
            \end{tabular}
        \end{center}

\subsection*{Table A V: Parameters for GEO -- GEO links}

        \begin{center}
            \begin{tabular}{|l|c|l|}\hline
                link distance       & $L$       & $40~000\,\mbox{km}$                  \\\hline
                aperture            & $D_T=D_R$ & $20\;\mbox{to}\;30\,\mbox{cm}$                   \\\hline
                wavelength          & $\lambda$ & $800\,\mbox{nm}$, $1550\,\mbox{nm}$ \\\hline
                telescope transmission factor & $T_T=T_R$ & $0.8$                                \\\hline
                pointing loss       & $L_p$     & $0.2$                                \\\hline
            \end{tabular}
        \end{center}


\section*{Acknowledgment}
This paper evolved from a project supported by the European Space
Agency under ESTEC/Contract No. 16358/02/NL/SFe, "Quantum
Communications in Space". We wish to thank Josep Maria Perdigues
Armengol for monitoring and supporting this work. We further
acknowledge support by the Austrian Science Foundation (FWF) and
the Alexander von Humboldt Foundation.
\newpage

\bibliographystyle{IEEEtran}
\bibliography{biblio}
\end{document}